\documentclass[12pt, onecolumn]{IEEEtran}
\usepackage{amssymb}
\usepackage{amsmath}
\usepackage{float}
\usepackage{algorithm}
\usepackage{algpseudocode}
\usepackage{times,amsmath,epsfig}
\usepackage[latin1]{inputenc}
\usepackage{fancyhdr}
\usepackage{amsfonts}
\usepackage{array}
\usepackage{graphicx}
\usepackage{epstopdf}
\usepackage{url}
\usepackage{subfigure} %¿É²åÈë×ÓÍ¼
\usepackage{bm}
\usepackage{breqn}
\usepackage{xcolor}
\usepackage{soul}
\usepackage{picinpar,graphicx}
\usepackage{cite}
\usepackage{multirow}
\ifCLASSINFOpdf
\else
\fi
\usepackage{amsmath}

\usepackage{graphicx}
\usepackage[justification=centering]{caption}
\usepackage{amsfonts}

\IEEEoverridecommandlockouts

\begin{document}
\begin{sloppypar}
\title{Intelligent Reflecting Surface assisted Integrated Sensing and Communication System}

\author{Zhiqing Wei, Xinyi Yang, Chunwei Meng, Xiaoyu Yang, \\
Kaifeng Han, Chen Qiu, Huici Wu
\thanks{Zhiqing Wei, Chunwei Meng, Xiaoyu Yang and Huici Wu are with Beijing University of Posts and Telecommunications, Beijing, China 100876 (email: \{weizhiqing, mengchunwei, xiaoyu.yang, dailywu\}@bupt.edu.cn), Xinyi Yang is with China Telecom Cloud Technology Co.,Ltd., Beijing, China (e-mail: yangxy18@chinatelecom), Kaifeng Han is with China Academy of Information and Communications Technology, Beijing, China (e-mail: hankaifeng@caict.ac.cn), Chen Qiu is with Peng Cheng Laboratory, Shenzhen, China (email: qiuch@pcl.ac.cn). \emph{Correspondence author: Zhiqing Wei.}}}
\maketitle

\begin{abstract}
High-speed communication and accurate sensing are of vital importance for future transportation system. Integrated sensing and communication (ISAC) system has the advantages of high spectrum efficiency and low hardware cost, satisfying the requirements of sensing and communication. Therefore, ISAC is considered to be a promising technology in the future transportation system. However, due to the low transmit power of signal and the influence of harsh transmission environment on radar sensing, the signal to noise ratio (SNR) at the radar receiver is low, which affects the sensing performance. This paper introduces the intelligent reflecting surface (IRS) into ISAC system. With IRS composed of $M$ sub-surfaces implemented on the surface of the target. The SNR at the radar receiver is 20lg($M$) times larger than the scheme without IRS. Correspondingly, radar detection probability is significantly improved, and Cramer-Rao Lower Bound (CRLB) for ranging and velocity estimation is reduced. This paper proves the efficiency of IRS enabled ISAC system, which motivates the implementation of IRS to enhance the sensing capability in ISAC system.
\end{abstract}

\begin{keywords}
Intelligent Reflecting Surface, Integrated Sensing and Communication, Detection Probability, Cramer-Rao Lower Bound
\end{keywords}
\IEEEpeerreviewmaketitle
%\keywords{Intelligent Reflecting Surface, Integrated Sensing and Communication, Detection Probability, Cramer-Rao Lower Bound}

\section{Introduction}

Driven by the development of safe and efficient transportation system, the vehicles require not only accurate sensing capabilities, but also high-speed data transmission for information interaction. Integrated Sensing and Communication (ISAC) technology integrates radar sensing and communication into one system, which reduces the hardware costs, improves the spectrum utilization and power efficiency \cite{1}. In addition, ISAC reduces the interference between sensing and communication by exchanging and sharing information between them \cite{2}. Therefore, ISAC is one of the potential key technologies to ensure high-speed communication and accurate sensing in the future transportation system. However, ISAC system applies low-power communication signal to realize radar sensing. The signal will scatter at the target, resulting in low signal to noise ratio (SNR) at the radar receiver, which limits the detection range and detection probability of radar. Therefore, improving the SNR at the radar receiver is the key to improve the performance of radar sensing.

In order to improve the SNR at radar receiver, methods such as relay amplification and forwarding \cite{3}, coherent integration \cite{4}, power control and beam diversity \cite{5} have been proposed. Relay amplification and forwarding can overcome the propagation of noise in the channel by re-encoding, amplification and forwarding the received signal. Coherent integration method improves the SNR since the phase fluctuation can be well compensated. However, long coherent integration time is needed, which makes it difficult to satisfy the requirements of fast sensing. By optimizing power control and beam allocation, the beam of radar reaches high magnitude so as to improve SNR at radar receiver. However, the computational complexity of this method is high.

In ISAC system, the above signal processing methods have limitations such as incompatibility with existing communication systems, high complexity, and so on. Therefore, a signal-processing-independent method is needed to avoid these limitations. Furthermore, the signal processing methods can be simultaneously applied to improve the sensing performance. Recently, intelligent reflecting surface (IRS) has been proposed in the field of communication. It is composed of passive reflecting elements that can adjust the reflection coefficient independently \cite{6}. In this paper, IRS assisted ISAC system is proposed, which could be used in vehicular sensing and communication. As illustrated in the case of Fig. 1, the IRS is deployed at the rear of the front vehicle to improve the SNR of echo signal at the rear vehicle. By adjusting the phase of reflected signal, IRS maximizes the strength of the echo signal, which improves the performance of radar sensing in ISAC system. The contributions of the paper are as follows. Firstly, the IRS realizes the superposition of echo signals through multiple passive reflecting sub-surfaces to obtain the echo signal with high power. Secondly, we proved that IRS increases the detection probability and reduces the CRLB of radar sensing. Finally, by optimizing the echo beam of the IRS, the SNR of the echo signal is maximized.

The rest of this paper is organized as follows. The system model is presented in section \uppercase\expandafter{\romannumeral2}. Section \uppercase\expandafter{\romannumeral3} derives the influence of IRS on radar echo power and CRLB for distance and velocity estimation, respectively. Section \uppercase\expandafter{\romannumeral4} presents the numerical results. In section \uppercase\expandafter{\romannumeral5}, the conclusions are presented.
\begin{figure}[h]
\centering
\includegraphics[width=9cm,height=3cm]{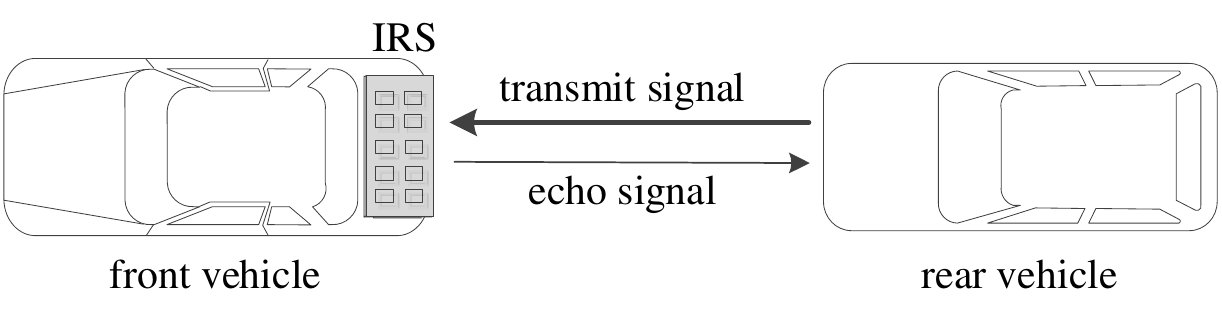}
\caption{IRS assisted ISAC system}
\end{figure}

\section{System Model}
The transmitter uses a single antenna for transmitting and receiving, IRS has $M$ sub-surfaces, as shown in Fig.1. The IRS is deployed at the rear of front vehicle, the rear vehicle transmits ISAC signal to the front vehicle, and obtains the velocity and distance information of the front vehicle according to the echo signal. Using advanced self-interference mitigation techniques, it is assumed that the IRS and the radar receiver  have perfect self-interference mitigation. The orthogonal frequency division multiplexing (OFDM) based ISAC signal is applied. The OFDM signal is expressed as \cite{1}
\begin{align}
x(t)=\sum_{\mu=0}^{N_{sym}-1}\sum_{n=0}^{N_c-1}A(\mu,n)d_{Tx}(\mu,n)e^{j2{\pi}f_nt}\rm rect(\frac{\textit{t}-{\mu}\textit{T}}{\textit{T}}),
\end{align}
where $N_{sym}$ represents the number of OFDM symbols, $N_c$ represents the number of subcarriers of an OFDM symbol. $\mu$ and $n$ represent symbol index and subcarrier index respectively, $d_{Tx}(\mu,n)$ represents the user data symbols, which are not related to any particular encoding. $A(\mu,n)$ represents the complex modulation symbols, including amplitude and phase shift. $f_n$ represents the carrier frequency. $\rm rect(\frac{\textit{t}-{\mu}\textit{T}}{\textit{T}})$ represents the baseband pulse shape, where $T$ represents the duration of OFDM symbol.

The transmit OFDM signal is reflected by the IRS with additional amplitude and phase information and received by the receiver together with delay and Doppler shift. The signal at the receiver is expressed as
\begin{align}
y(t)=x(t-\tau)e^{j2{\pi}f_dt}H_M,
\end{align}
where $\tau=\frac{2R}{C_0}$ represents the time delay, $R$ represents the distance between the transmitter and the target, $c_0$ represents the speed of light. $f_D=\frac{2\nu f_c}{c_0}$ represents the Doppler shift, $\nu$ represents the velocity of the target, $f_c$ represents the carrier frequency. $H_M$ represents the sensing channel from transmitter to receiver, specifically expressed as
\begin{align}
H_M=\emph{G}_t^T \mathbf{\Psi}_M \emph{G}_r,
\end{align}
where $\emph{G}_r\in\mathbb{C}^{M\times 1}$ represents the channel from transmitter to IRS, $\emph{G}_t\in\mathbb{C}^{M\times 1}$ represents the channel from IRS to receiver, $\mathbf{\Psi}_M\in\mathbb{C}^{M\times M}$ represents a diagonal matrix with phase shift $e^{j\mathbf{\Psi}_m}$ and amplitude shift $\alpha_m$. $m$ representing the $m$-th diagonal element of the matrix, $\alpha_m e^{j\mathbf{\Psi}_m}$  constitutes the reflection coefficient of the $m$-th sub-surface of the IRS. The reflection coefficient characterizes the equivalent interaction of the $m$-th sub-surface with the incident signal \cite{8}.

IRS improves the sensing performance of the ISAC system by changing the amplitude and phase of each sub-surface. The radar needs to locate the target by extracting the phase information of the echo signal. In the above echo signal model, the IRS will change the phase of the echo signal. Then, its impact on the sensing performance is analyzed.

First, the complete received signal is expressed as
\begin{align}
y(t)=\sum_{\mu=0}^{N_{sym}-1}\sum_{n=0}^{N_c-1}H_M A(\mu,n)d_{Tx}(\mu,n)\\e^{j2{\pi}f_n(t-\frac{2R}{c_0})}e^{j2\pi\frac{2\nu f_c}{c_0}t}\rm rect(\frac{\textit{t}-{\mu}\textit{T}-\frac{2\textit{R}}{c_0}}{\textit{T}}).
\end{align}

When the protection time interval is properly selected, the rect-function can be neglected. For a fixed OFDM symbol index $\mu$, selecting a different subcarrier has the same effect on the Doppler frequency shift. So the velocity of the target is related to the position of the OFDM symbol, instead of the subcarrier position on a symbol, which means that we can estimate the target velocity from the frequency domain. Similarly, for a fixed subcarrier index $n$, selecting a different OFDM symbol has the same effect on time delay. The range of the target is related to the position of the subcarrier, which means that we can estimate the distance from the time domain.

Considering the influence of OFDM symbol index and subcarrier index on the echo signal, the received data symbols $d_{Rx}(\mu,n)$ can be expressed as
\begin{align}
d_{Rx}(\mu,n)=A(\mu,n)H_Md_{Tx}(\mu,n)e^{-j2{\pi}n\Delta f\frac{2R}{c_0}}e^{j2{\pi}\mu T\frac{2\nu f_c}{c_0}},
\end{align}
where $\Delta f$ represents subcarrier spacing. Converting the transmitted and received data symbols into transmitted matrices $(\mathbf{D}_{Rx})_{\mu,n}$ and received matrices $(\mathbf{D}_{Tx})_{\mu,n}$, where each row of the matrix represents a subcarrier and each column represents an OFDM symbol. To get the velocity and distance information of the target, the transmitted data symbols need to be removed from the received data symbols by an element-wise complex division. Finally, a two-dimensional matrix $\mathbf{D}_{div}$ is obtained, where each element $(\mathbf{D}_{div})_{\mu,n}$ can be represented as
\begin{align}
(\mathbf{D}_{div})_{\mu,n}=\frac{(\mathbf{D}_{Rx})_{\mu,n}}{(\mathbf{D}_{Tx})_{\mu,n}}=A(\mu,n)H_Me^{-j2{\pi}n\Delta f\frac{2R}{c_0}}e^{j2{\pi}\mu T\frac{2\nu f_c}{c_0}}.
\end{align}

Based on the single-antenna transmission and single-antenna reception vehicle network scenario studied in this paper, the sensing performance of echo signals is enhanced by optimizing the amplitude and phase of each subsurface of the IRS. The change of the IRS reflection coefficient is independent of the symbol index and subcarrier index of OFDM. Assume that the sensing channel is constant during the time from signal transmission to signal reception. By designing the reflection coefficient of the IRS, eventually constant $H_M$ over a very short sensing time is obtained, together with $A(\mu,n)$ forming a complex number.

In (7), the distance information of the target is converted into a linear phase shift along the frequency domain. Applying inverse discrete Fourier transform on $e^{-j2{\pi}n\Delta f\frac{2R}{c_0}}$, the time domain response function is obtained. The distance between the radar and the target can be estimated by obtaining the index position corresponding to the peak value of the time domain response function. Similarly, applying discrete Fourier transform on $e^{j2{\pi}\mu T\frac{2\nu f_c}{c_0}}$, the frequency domain response function is obtained. The velocity between the radar and the target can be estimated by obtaining the index position corresponding to the peak value of the frequency domain response function. Since $A(\mu,n)H_M$ is a complex number, which will not change the maximum value, the application of IRS will not cause errors in range and velocity estimation.

\section{IRS Assisted Radar Detection}
\subsection{Transmit power of radar with IRS}
The IRS is covered on the surface of target to be detected. The echo signal from the IRS is received at the radar receiver. These manually controlled meta-surfaces can be regarded as ideal phase adjusters. Therefore, the multipath components can be superposed coherently to maximize the SNR at radar receiver.

The maximum transmit power density of radar is
\begin{align}
 Q_t=\frac{P_t G}{4\pi R^2},
\end{align}
where $P_t$ is the transmit power, $G$ is the transmit antenna gain, $R$ is the distance between radar and target. Assuming that radar cross section (RCS) is $\sigma$, the power of echo signal is
\begin{align}
P_b=\frac{P_t G \sigma}{4 \pi}{\left|\sum_{m=1}^M\frac{\beta_{m} \alpha_{m}e^{j\left(\theta_{m}-\phi_{m}\right)}}{R_m}\right|}^2.
\end{align}

The IRS is made up of $M$ reconfigurable sub-surfaces, each of which is capable of tuning the angle of reflection according to the Snell's law and adjusting the phase of the reflected signal independent of the other sub-surfaces \cite{7}. The index $m$ refers to the $m$-th reconfigurable sub-surface of IRS. $R_m$ is the distance between radar and each sub-surface. Because the distance between the sub-surfaces is much smaller than that between the transmitter and IRS, we denote $R_m=R$ for all $m$. Then, the power density of the echo signal of radar is as follows.
\begin{align}
Q_{b}=P_{t} G \sigma \frac{\left|\sum_{m=1}^{M} \beta_{m} \alpha_{m}e^{j\left(\theta_{m}-\phi_{m}\right)}\right|^{2}}{\left(4 \pi R^{2}\right)^{2}}.
\end{align}

The received power of echo signal is
\begin{align}
P_r=P_t G^2 \sigma {\lambda}^2 \frac{\left|\sum_{m=1}^{M} \beta_{m} \alpha_m e^{j\left(\theta_{m}-\phi_{m}\right)}\right|^{2}}{(4\pi)^3 R^4 L},
\end{align}
where $\lambda$ indicates the wavelength of signal, and $L$ indicates the path loss factor.

In order to reveal the impact of IRS on the received power of radar, we consider the ideal situation where the backscattering of each element is coherently superimposed. The total amplitude is equal to the sum of the amplitudes of all scattering elements, and the power is equal to the square of the sum of the amplitudes. The power of the echo signal under ideal conditions is
\begin{align}
P_r=\frac{P_tG^2 \sigma {\lambda}^2  {\left|\sum_{m=1}^{M} \beta_{m} \alpha_m\right|^{2}}}{(4\pi)^3 R^4 L}.
\end{align}

The SNR of the echo signal at the receiver is

\begin{align}
\gamma=\frac{P_t G^2 \sigma {\lambda}^2 \left|\sum_{m=1}^{M} \beta_m \alpha_m\right|^2}{(4\pi)^3 R^4 L N_0}.
\end{align}

It is worth noting that when $M=1$, IRS will not bring additional power gain. When there are multiple sub-surfaces, the power of echo signal is enhanced.

According to (10), when $\beta_m=\alpha_m=1$, SNR is proportional to the square of the number of sub-surfaces $M$. Compared with the scheme without IRS, the SNR at the radar receiver has increased by 20lg($M$) times.
\subsection{Radar Detection Probability and CRLB}
One of the performance indicators of radar sensing is the detection probability. In radar sensing, the detection probability and false alarm probability depend on the threshold of detection. The principle of Constant False Alarm Rate (CFAR) algorithm is to automatically adjust the detection threshold according to the strength of interference to keep the false alarm probability constant. By comparing the amplitude value of the signal to be detected with the detection threshold \cite{9}, the existence of target is detected.

The amplitude and phase of the echo signal reflected from the target are unknown, and the envelope of narrow-band white Gaussian noise in the channel follows Rayleigh distribution. The square-law detector is used at the radar receiver. According to CFAR algorithm \cite{9}, the average false alarm probability $P_{fa}$ and average detection probability $P_d$ can be derived as
\begin{align}
P_{fa}=(1+\frac{\alpha}{N})^{-N},
\end{align}
\begin{align}
P_d=(1+\frac{\alpha}{N(1+\gamma)})^{-N},
\end{align}
where $\alpha$ represents the scale factor, $\gamma$ represents the SNR at radar receiver. As the number of interference signals $N$ increases, $P_{fa}$ and $P_d$ are only related to the SNR at radar receiver. Finally, the radar detection probability is
\begin{align}
P_d={P_{fa}}^{\frac{1}{1+\gamma}}.
\end{align}
Substituting (10) into (13), since the IRS improves the SNR of the echo signal, the target has larger probability to be detected by radar.

In order to evaluate the performance of velocity and distance estimation, we use Cramer-Rao Lower Bound (CRLB) to determine the minimum variance of the unbiased estimation \cite{10}.
%Specifically, we first characterize the Fisher matrix to derive CRLB theoretically, where each element can be represented as
%\begin{align}
%{{[I( \boldsymbol{B})]}_{ij}}=\operatorname{Re}\{{{\sum\limits_{n=0}^{N-1}{\left[ \frac{\partial s[n;\boldsymbol{B} ]}{\partial {{\boldsymbol{B} }_{i}}} \right]}}^{*}}\left[ \frac{\partial s[n;\boldsymbol{B} ]}{\partial {{\boldsymbol{B} }_{j}}} \right]\},
%\end{align}
%where $\boldsymbol{B}$ stands for ternary parameter vector ${{\left[ \left| {{H}_{M}} \right|,f,\left( \theta -\phi  \right) \right]}^{T}}$, $f$ represents signal frequency, $s[n;\boldsymbol{B}]$ can be expressed as
%\begin{align}
%s\left[ n;\boldsymbol{B}  \right]=\left| {{H}_{M}} \right|A\left( \mu ,n \right){{e}^{j\left( \theta -\phi  \right)}}{{e}^{j2\pi nf}}.
%\end{align}
To obtain the CRLB, Fisher matrix needs to be obtained first. According to the first-order partial derivative of the estimated parameters, the Fisher matrix can be obtained as
\begin{align}
\left[ \begin{matrix}
   A{{\left( \mu ,n \right)}^{2}}{{N}_{c}} & 0 & 0  \\
   0 & {{\left( 2\pi \left| {{H}_{M}} \right|A\left( \mu ,n \right) \right)}^{2}}{{\Gamma}} & \pi {{\left( \left| {{H}_{M}} \right|A\left( \mu ,n \right) \right)}^{2}}{{\Upsilon}}  \\
   0 & \pi {{\left( \left| {{H}_{M}} \right|A\left( \mu ,n \right) \right)}^{2}}{{\Upsilon}} & {{\left( \left| {{H}_{M}} \right|A\left( \mu ,n \right) \right)}^{2}}{{N}_{c}}  \\
\end{matrix} \right],
\end{align}
where ${{\Upsilon}}={{N}_{c}}({{N}_{c}}-1)$, ${{\Gamma}}=\frac{{{\Upsilon }}(2{{N}_{c}}-1)}{6}$, $N_c$ represents the number of OFDM subcarriers. Then, based on the inverse matrix of Fisher matrix, using the similar techniques as \cite{10}
, the CRLB of phase estimation can be given by
\begin{align}
CRLB(f)=\frac{6}{^{{}}{{\left( 2\pi \left| {{H}_{M}} \right|A\left( \mu ,n \right) \right)}^{2}}{{N}_{c}}(N_{c}^{2}-1)}.
\end{align}
%Substituting the OFDM symbol into (18) [1],
The CRLB of distance estimation can be obtained as follows.
\begin{align}
CRLB(R)=\frac{6c_0^2}{(4\pi\Delta f)^2|H_M|^2A(\mu,n)^2{N_c}(N_c^2-1)}.
\end{align}
In practical applications, each symbol of the signal matrix at the receiver has different Doppler frequency domains, which makes it complicated to calculate the CRLB of distance estimation in a whole frame. Therefore, the mean value can be used to determine a simpler lower bound, which can be expressed as
\begin{align}
CRLB(R)=\frac{6c_0^2}{(4\pi\Delta f)^2|H_M|^2A(\mu,n)^2N_{sym}{N_c}(N_c^2-1)},
\end{align}
where $N_{sym}$ represents the number of OFDM symbols. Due to $H_M=\sum_{m=0}^{M-1}\beta_me^{-j\phi_m}\alpha_me^{j\theta_m}$, the optimal solution of $\theta_m$ that maximizes the SNR is $\theta_m$ = $\phi_m$ for $m$ = 1,...,$M$.

Similarly, taking the average value, the CRLB the velocity estimation is obtained as follows.
\begin{align}
CRLB(\upsilon)=\frac{6c_0^2}{(4\pi Tf_c)^2|H_M|^2A(\mu,n)^2N_{sym}N_c(N_{sym}^2-1)}.
\end{align}

Because the IRS can be used to maximize the SNR by appropriately designing the IRS reflection coefficient. The multipath components at the radar receiver are superimposed coherently, and the SNR at radar receiver is maximized through additional phase shift generated by IRS. With $\boldsymbol{Z} =\left\{\theta_{1}, \theta_{2}, \cdots, \theta_{M}\right\}$ representing the designed phase shift vector generated at IRS, an objective function can be expressed as follows.
\begin{align}
\underset{\boldsymbol{Z}}{\text { Maximize }} \operatorname{\gamma}(\boldsymbol{Z})=\frac{P_{t} G^{2} \lambda^{2}\left|\sum_{m=1}^{M} \beta_{m} r_me^{j\left(\theta_{m}-\phi_{m}\right)}\right|^{2}}{N_{0}(4 \pi)^{3} R^{4} L}.
\end{align}

The optimal SNR in (22) is obtained by using the stochastic gradient descent algorithm \cite{11}. Substituting the optimal solution obtained by (22) into (13), (14) and (15), it can be discovered that the increase in the power of the echo signal improves the SNR at radar receiver, thus maximizing the performance of radar detection probability and CRLB.
\section{Simulation Results}
In this section, we provide simulation results to verify the effectiveness of the proposed IRS assisted ISAC system. The parameters of OFDM signal are $N_{sym}$=12, $N_c$=512, $\Delta f$=30 kHz, $T$=33.3 $\mu s$. The transmit signal power is 20 W and the noise power is 0.8 W. The modulation scheme is 16QAM. The following results are obtained by averaging over 1000 Monte Carlo results.

Fig. 2 illustrates the relation between the reflection coefficient of IRS and the SNR of echo signal. For comparison, the influence of IRS on SNR is illustrated for the schemes with optimal IRS coefficients, with random IRS coefficient, and without IRS. As illustrated in Fig. 2, the optimal IRS reflection coefficients obtained by the gradient descent algorithm improve the SNR significantly. In addition, the SNR for all these cases increases with the increase of the number of sub-surfaces of IRS, because more sub-surfaces can provide richer multipath information.
\begin{figure}[h]
\centering
\includegraphics[width=10cm,height=7cm]{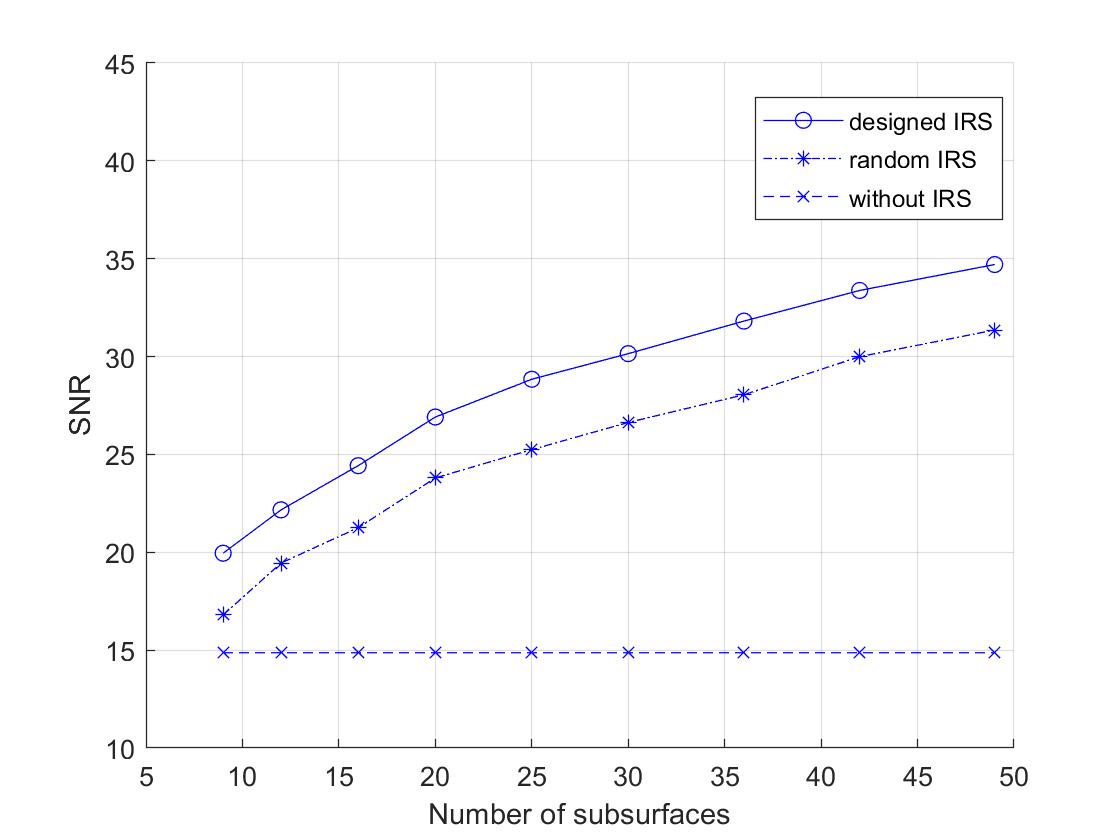} %.eps
\caption{The relation between the SNR of echo signal and the number of sub-surfaces}
\end{figure}

Fig. 3 presents the relation between the power of echo signal and the SNR. Firstly, it is observed that the power of echo signal has a linear relationship with the SNR, and the power of echo signal shows an upward trend with the increase of the SNR.
Secondly, compared with the scheme without IRS, the power of echo signal with IRS is significantly improved. In addition, the power of echo signal is proportional to the number of sub-surfaces. Thus, the sensing performance of ISAC system is improved when the number of sub-surfaces is increased.
\begin{figure}[h]
\centering
\includegraphics[width=10cm,height=7cm]{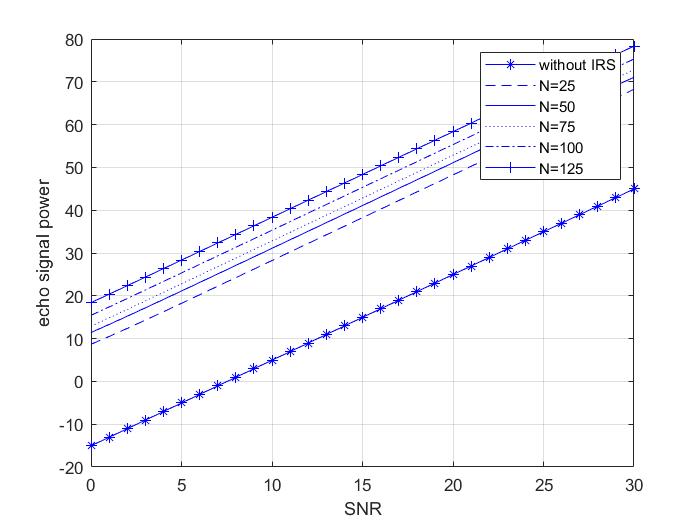}
\caption{The relation between the power of echo signal and the SNR}
\end{figure}
%The relation between the power of echo signal and the number of sub-surfaces

The relation between the number of sub-surfaces and detection probability under different $P_{fa}$ is shown in Fig. 4. When the number of sub-surfaces is the same, the $P_d$ increases when the $P_{fa}$ increases. In practical applications, it is more desirable to obtain a small $P_{fa}$ and a large $P_d$. When the $P_{fa}$ is the same, the $P_d$ not only increases with the increase of the number of sub-surfaces, but also is higher than the ISAC system without IRS. Therefore, the application of IRS can achieve a larger $P_d$ under the condition of a smaller $P_{fa}$.
\begin{figure}[h]
\centering
\includegraphics[width=10cm,height=7cm]{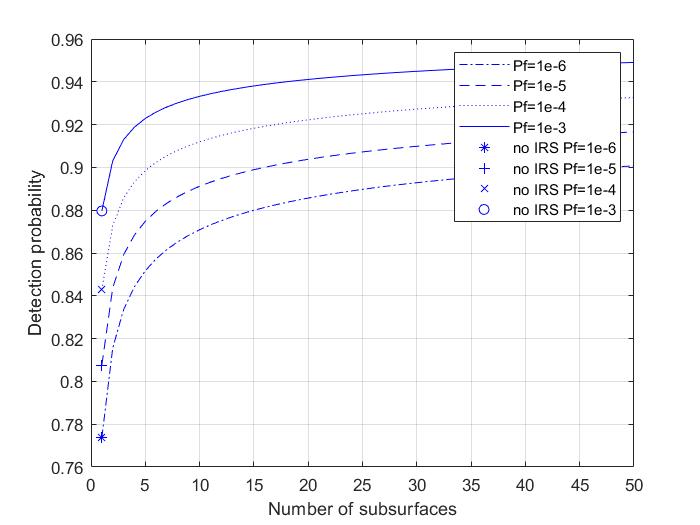}
\caption{The relation between the number of sub-surfaces and detection probability under different false alarm probability}
\end{figure}

%The performance of range estimation in terms of MMSE is shown in Fig. 5. It is observed that the CRLB of radar range estimation behaves differently under different number of sub-surfaces. Firstly, compared with the scheme without IRS, the application of IRS reduces the CRLB of ranging. Secondly, as the number of sub-surfaces increases, CRLB decreases. Overall, the IRS-assisted ISAC system provides better MMSE performance.
Fig. 5 shows the CRLB and minimum mean squared error (MMSE) of distance estimation. The range estimation CRLB with different number of sub-surfaces is compared with the range estimation MMSE without IRS and the range estimation CRLB without IRS. Firstly, compared with the range estimation scheme without IRS, the application of IRS reduces the CRLB of distance estimation. Secondly, as the number of sub-surfaces increases, CRLB decreases. Overall, the IRS-assisted ISAC system provides better MMSE performance.
%It is observed that the CRLB of radar range estimation behaves differently under different number of sub-surfaces.
\begin{figure}[h]
\centering
\includegraphics[width=10cm,height=7cm]{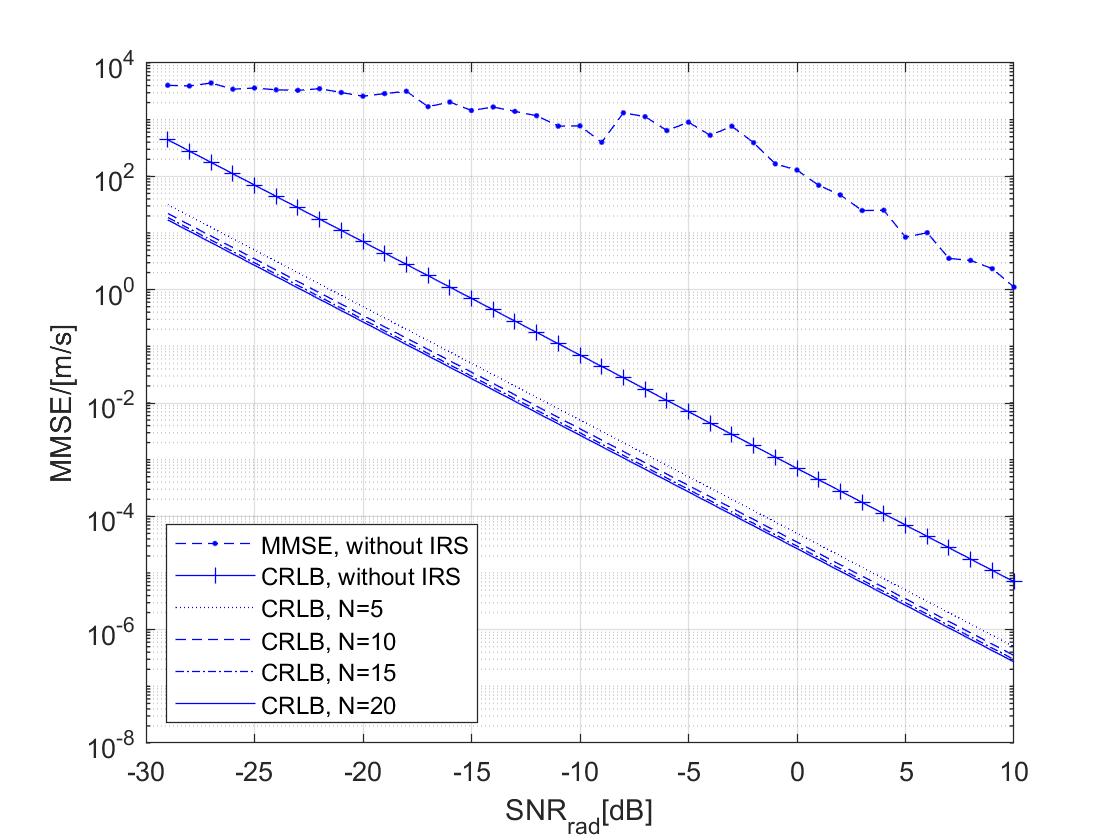}
\caption{The relation between the MMSE of the target range estimation and the SNR}
\end{figure}

\section{Conclution}
In this paper, we have investigated the IRS assisted ISAC system, which improves the power of echo signal, thus enhancing the sensing performance of ISAC system using low-power communication signal. It is discovered that SNR at the radar receiver is 20lg($M$) times larger compared with the scheme without IRS, with $M$ denoting the number of subsurfaces. The detection probability and CRLB of radar sensing are adopted as the performance metrics, which are derived and simulated. It is verified that the IRS improves the detection probability and reduces the CRLB. Besides, the increase of the number of subsurfaces further enhances these performance metrics. This paper proves the advantages and feasibility of IRS assisted ISAC system, which may motivates the application of IRS in the ISAC system to enhance the sensing performance.

\section{Acknowledgement}
This work was supported in part by the National Key Research and Development Program under Grant 2020YFA0711302, in part by the National Natural Science Foundation of China under Grant U21B2014, in part by the National Natural Science Foundation of China under Grant 62101293, China Postdoctoral Science Foundation under Grant 2021M701806, and the Major Key Project of PCL (PCL2021A15).

\bibliographystyle{IEEEtran}
\bibliography{ciations}

\end{sloppypar}
\end{document}